\documentclass[superscriptaddress,twocolumn,showpacs,preprintnumbers,amsmath,amssymb,amsfonts,aps,prb]{revtex4-1}
\usepackage{graphicx}
\usepackage{dcolumn}
\usepackage{bm}

\providecommand{\abs}[1]{\lvert#1\rvert}

\begin{document}
%
%
\title{Beam shifts and distribution functions}
\author{Andrea Aiello}
\email{andrea.aiello@mpl.mpg.de}
\affiliation{Max Planck Institute for the Science of Light, G\"{u}nther-Scharowsky-Stra{\ss}e 1/Bau 24, 91058 Erlangen, Germany}
\affiliation{Institute for Optics, Information and Photonics, Universit\"{a}t Erlangen-N\"{u}rnberg, Staudtstr. 7/B2, 91058 Erlangen, Germany}
\date{\today}
%
\begin{abstract}
When a beam of light is reflected by a smooth surface its behavior deviates from geometrical optics predictions. Such deviations are quantified by the so-called spatial and angular Goos-H\"{a}nchen (GH) and Imbert-Fedorov (IF) shifts of the reflected beam.
These shifts depend upon the shape of the incident beam, its polarization and on the material composition of the reflecting surface.
In this article we suggest a novel approach  that allows one to unambiguously isolate the beam-shape dependent aspects of GH and IF shifts.
We show that this separation is possible as a result of some
 universal features of shifted distribution functions which are presented and discussed.
\end{abstract}

\maketitle

%
\section{Introduction}
The reflection and refraction of light at the plane surface of a transparent medium, such as the still water of a placid lake, are well-known phenomena. As a matter of fact, the \emph{law of reflection} was already enunciated by Euclid of Alexandria (fl. $300$ BC) in his book \emph{Catoptrics}. \cite{HechtBook} Nowadays, extended expositions of this subject may be found in standard optics textbooks. \cite{JacksonBook,BandWBook}

Less well-known is the fact that when the light has the form of a beam with a finite transverse (i.e., orthogonal to the  direction of propagation) extent, the reflected beam appears shifted with respect to a ray traveling parallel to the  beam  but reflected according to the laws of \emph{geometrical optics}, as opposed to the laws of \emph{wave optics} obeyed by the finite beam. A cartoon-like representation of this phenomenon is presented in Fig. 1.
%
%
\begin{figure}[h!]
\begin{center}
\includegraphics[angle=0,width=8truecm]{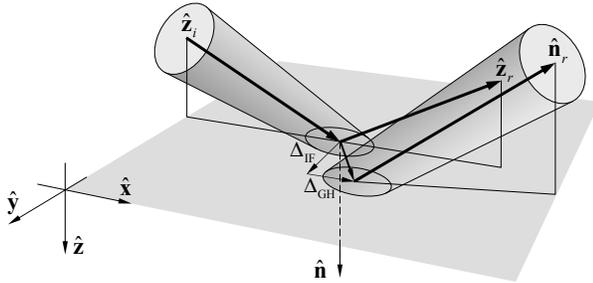}
\caption{\label{fig:1} Illustration of a beam-shift process. The \emph{incident} beam travels along with the ray represented by the unit vector $\hat{\mathbf{z}}_i$.  Reflection transforms $\hat{\mathbf{z}}_i$  according to the laws of geometrical optics into $\hat{\mathbf{z}}_r = \hat{\mathbf{z}}_i - \hat{\mathbf{n}}\left( \hat{\mathbf{n}} \cdot \hat{\mathbf{z}}_i \right) $, where $\hat{\mathbf{n}}$ is a unit vector normal to the reflecting surface. The \emph{reflected} beam travels parallel the unit vector $\hat{\mathbf{n}}_r$ which is both displaced and tilted with respect to $\hat{\mathbf{z}}_r$. $\Delta_\text{GH}$ and $\Delta_\text{IF}$ denote, respectively,  the projections onto the plane of the surface of the beam displacements parallel and orthogonal to the plane of incidence spanned by   $\hat{\mathbf{z}}_i$ and   $\hat{\mathbf{n}}$.}
\end{center}
\end{figure}
%
%

Such a shift may have both a spatial and an angular character, namely the reflected beam may be both \emph{displaced} and \emph{deflected} with respect to the geometrical ray. Depending on whether these spatial and angular shifts occur in either the plane of incidence or in a direction orthogonal to this plane, they are denoted as either Goos-H\"{a}nchen (GH) or Imbert-Fedorov (IF) shifts. Analogous shifts also occur for the refracted beam.

All these shifts are very small compared to the transverse extent of the beam, typically a few wavelengths  the spatial shifts, and from micro- to milli-radians  the angular ones.
Interestingly enough,   the first who conjectured that the center of a  beam of light should manifest a tiny spatial shift in the plane of incidence when reflected, was Isaac  Newton in the 17th century. \cite{NewtonOptiks}

Detailed theoretical derivations of both GH and IF shifts may be found in the current literature for surfaces of different types (dielectric, metallic, multilayered, et cetera), and for light beams of vary shapes (Hermite-Gauss, Laguerre-Gauss, Bessel, et cetera).  The interested reader may consult Refs. 5-8
 for early expositions and Refs. 9-15
for more accessible modern treatments of the subject.
From these studies it emerges that, given a specific reflecting surface,
beams of different shapes yield shifts of different extent. For example, for an air-to-glass interface,  a fundamental Gaussian beam produces qualitatively and quantitatively different GH and IF shifts with respect to a Laguerre-Gauss beam impinging upon the same surface. This could appear not so surprising because a Laguerre-Gauss beam carries an intrinsic orbital angular momentum while a fundamental Gaussian beam does not. \cite{Allen92} Thus, one might conclude
that orbital angular momentum (OAM) of light affects the physics of beam-shift phenomena.
More generally, since such OAM  is determined solely by the spatial profile of the beam,  one may wonder to what extent the beam  shape determines the values of both GH and IF shifts.

Unfortunately, prevailing discussions in the current literature tend to obscure the distinction between the beam-shape dependent and independent aspects of beam-shift phenomena  because conventional evaluations of GH and IF shifts  rest on the direct calculation of the centroid of the intensity distribution of the reflected beam. Such a distribution depends, as a whole, upon \emph{both} the shape of the beam \emph{and}  the material composition of the surface.   Thus,
 beam-shape dependent and independent characteristics of shift phenomena deceptively appear to be intrinsically tied.

In this article we present a novel way to describe the GH and IF shifts, with the aim to provide significantly new insights about these phenomena. To this end, a simple treatment that renders a clear separation between the  beam-shape dependent and independent features of beam-reflection processes is presented and developed. Such separation naturally leads to find some ``universal'' characteristics of the beam-shape dependent facets of the problem.

The rest of this paper is organized as follows: In Sec. II the conventional theory of GH and IF shifts is reviewed  and  expressions for the beam-shape independent parts of these shifts are explicitly given. In Sec. III and IV a new treatment based on the study of the functional shape of the beam is introduced and illustrated. Finally, in Sec. V, consequences of the preceding results are discussed.
\section{Goos-H\"{a}nchen and Imbert-Fedorov shifts}
Consider a well collimated monochromatic beam of light  of wavenumber $k$  propagating in the direction $\hat{\mathbf{z}}_i$. The geometric details are shown in Fig. 2 below.
%
%
\begin{figure}[hb!]
\begin{center}
\includegraphics[angle=0,width=8truecm]{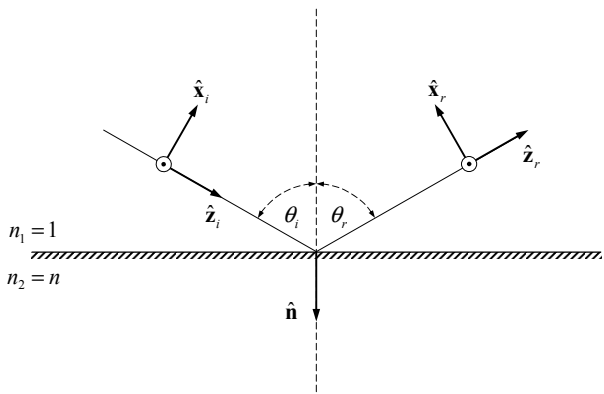}
\caption{\label{fig:2} Geometric representation of the system. The plane of incidence coincides with the plane of the figure.
The unit vector $\hat{\mathbf{n}}$ is normal to the interface between media $1$ (top) and $2$ (bottom). The unit vectors $\hat{\mathbf{y}}_i = \hat{\mathbf{y}}_r \equiv \hat{\mathbf{y}}$ (not shown in the figure), are perpendicular to the plane of incidence and directed towards the reader. The Snell law of reflection sets $\theta_i = \theta_r \equiv \theta$. The index of refraction $n_1$ of the first medium is conventionally chosen equal to $1$, and the index of refraction $n_2$ of the second medium is  $n$.}
\end{center}
\end{figure}
%
%

The position vector $\mathbf{r}$ has  Cartesian  coordinates ${ \mathbf{r} = (x' \, \hat{\mathbf{x}}_i +  y' \, \hat{\mathbf{y}}_i +  z' \, \hat{\mathbf{z}}_i)/k}$ and ${\mathbf{r} = (x \, \hat{\mathbf{x}}_r +  y \, \hat{\mathbf{y}}_r +  z \, \hat{\mathbf{z}}_r)/k}$ with respect to the reference frames $\{ \hat{\mathbf{x}}_i , \hat{\mathbf{y}}_i , \hat{\mathbf{z}}_i \}$ and $\{ \hat{\mathbf{x}}_r , \hat{\mathbf{y}}_r , \hat{\mathbf{z}}_r \}$ attached to the incident and to the reflected beam, respectively. Hereafter all the distances $x',y', \ldots$ are dimensionless. The corresponding dimensional quantities are simply equal to $x'/k, \, y'/k, $ et cetera.
The real electric field associated with such a beam may be written as
\begin{align}
\mathbf{E}'(x',y',z',t) = \text{Re} \left[ \mathbf{A}' (x',y',z') \exp\left(- i k \, c \, t \right)\right], \label{EfieldInc}
\end{align}
where $c$ denotes the speed of light in vacuum and
\begin{align}
\mathbf{A}' (x',y',z') = \left( a_{\, \parallel} \, \hat{\mathbf{x}}_i + a_\perp \, \hat{\mathbf{y}}_i \right) f(x',y',z'), \label{AnaSigInc}
\end{align}
is the so-called \emph{complex analytic signal} \cite{MandelBook} associated to the real signal $\mathbf{E}'(x',y',z',t)$.
In Eq. \eqref{AnaSigInc} and hereafter, subscripts ``$\parallel$'' and ``$\perp$'' refers, respectively, to the directions parallel and perpendicular to the plane of incidence. In optics textbooks such directions are often denoted with the letters ``$p$'' and ``$s$'', respectively.
The two complex amplitudes $a_{\, \parallel}$ and $a_{\, \perp}$
define the polarization of the beam and the complex scalar function $f(x',y',z')$ fixes the spatial shape of the beam. For example, $f(x',y',z')$ may either  be a Hermite-Gauss function or a Laguerre-Gauss function, both types being widely utilized in  optics. \cite{SiegmanBook}

When the beam is reflected its analytic signal $\mathbf{A}' (x',y',z')$  changes into $\mathbf{A} (x,y,z)$, where \cite{Note}
\begin{align}
\mathbf{A} (x,y,z) = & \;  a_{\, \parallel} r_{\parallel}(\theta) f(-x + X_{\, \parallel} ,y - Y_{\, \parallel},z) \, \hat{\mathbf{x}}_r \nonumber \\
& \; + a_\perp r_{ \perp}(\theta) f(-x + X_{ \perp} ,y - Y_{\perp},z)\, \hat{\mathbf{y}}_r  , \label{AnaSigRef}
\end{align}
and $r_{\parallel}(\theta), \, r_{\perp}(\theta)$ are the  Fresnel reflection coefficients evaluated at the incident angle $\theta_i = \theta$. The  four dimensionless \emph{complex} shifts $X_{\, \parallel}, Y_{\, \parallel}$ and  $X_\perp , Y_\perp$ are given by: \cite{AielloPra2010}
\begin{subequations}\label{ComplexShifts}
\begin{align}
X_{\, \parallel} = & \; -i \frac{\partial \ln r_{\parallel}}{\partial \theta},   & \;
Y_{\, \parallel} &   =  i \frac{a_{ \perp}}{a_{\, \parallel}}\left( 1 + \frac{r_{ \perp}}{r_{\parallel}}\right) \cot \theta,  \label{Yp} \\
X_{ \perp} = & \; -i \frac{\partial \ln r_{ \perp}}{\partial \theta},  & \;
Y_{ \perp} &   =  -i \frac{a_{\, \parallel}}{a_{ \perp}}\left( 1 + \frac{r_{ \parallel}}{r_{ \perp}}\right) \cot \theta. \label{Yo}
\end{align}
\end{subequations}
These shifts depend upon the polarization of the incident beam via the complex amplitudes $a_{\, \parallel}$ and $a_{\, \perp}$, and upon the characteristics of the surface via the (either real or complex) Fresnel coefficients $r_{\parallel}$ and $r_{\perp}$. However, they are \emph{completely} independent from the shape of the beam $f(x',y',z')$. The reader should notice and keep in mind this important fact.  Equations \eqref{AnaSigRef} and \eqref{ComplexShifts} are valid under the assumption that the reflection does not strongly distort the beam. Such deformation occurs, for example, when the beam impinges at the Brewster angle \cite{HechtBook} for which one has $r_\parallel =0$ and $Y_\parallel$ becomes ill-defined; in this case a special treatment is required. \cite{AielloOL2009}

For monochromatic light the intensity of the reflected beam is given by: \cite{HechtBook}
\begin{align}
I (x,y,z) = & \; \frac{1}{2} \left(\frac{\varepsilon_0}{\mu_0}\right)^{1/2} \mathbf{A}^* \cdot \mathbf{A}  \nonumber \\
= & \; \frac{1}{2} \left(\frac{\varepsilon_0}{\mu_0}\right)^{1/2} \left(I_{\parallel} + I_{\perp} \right), \label{IntRef}
\end{align}
where $\varepsilon_0$ and $\mu_0$ are the vacuum permittivity and permeability, \cite{JacksonBook} respectively, and
\begin{subequations}\label{IntPO}
\begin{align}
I_\parallel = & \; \left|a_{\, \parallel} \, r_{\parallel} \right|^2 \, \left|f(-x + X_{\, \parallel} ,y - Y_{\, \parallel},z) \right|^2,  \label{IntPOa} \\
I_\perp = & \; \bigl|a_{ \perp} \, r_{\perp} \bigr|^2 \, \bigl| f(-x + X_{\perp} ,y - Y_{ \perp},z) \bigr|^2, \label{IntPOb}
\end{align}
\end{subequations}
are the intensity distribution functions for the reflected light with polarization parallel and perpendicular to the plane of incidence, respectively.

The center of the reflected beam may be evaluated as the centroid  in the $xy$ plane of the total intensity distribution $I_\parallel + I_\perp$. The position of such centroid with respect to the origin $x=0, \, y=0$ on the plane $z = \text{const.}$, defines both the spatial and angular GH and IF shifts  in the following way.
Given the Cartesian components $x_1 =x$ and $x_2 = y$ of the centroid of the beam defined as:
\begin{align}
\langle x_\alpha \rangle(z) = \frac{\displaystyle{\int x_\alpha \left( I_\parallel + I_\perp \right) \; d x \, d y}}{\displaystyle{\int  \left( I_\parallel + I_\perp \right) \; d x \, d y} }, \qquad (\alpha = 1,2),\label{AveX}
\end{align}
 then the spatial ($\Delta$) and angular ($\Theta$) GH and IF shifts are conventionally \emph{defined} as:
\begin{subequations} \label{GHandIFShifts}
\begin{align}
\Delta_\text{GH} & =  \langle x \rangle (0), \qquad & \Theta_\text{GH} & =  \frac{ \partial \, \langle x \rangle(z)}{\partial z},
 \label{GHShifts} \\
\Delta_\text{IF} & =  \langle y \rangle (0), \qquad & \Theta_\text{IF} & =  \frac{ \partial \, \langle y \rangle(z)}{\partial z}.
 \label{IFShifts}
\end{align}
\end{subequations}
Note that in Eq. \eqref{AveX} and hereafter the  integration symbol ``$\int \,$'' \emph{always} denotes definite double integrals  over the whole $xy$ plane with $z$ kept constant. Moreover, the fact  that $\Theta_\text{GH}$ and $\Theta_\text{IF}$ are strictly independent from $z$ (see, e. g.,  Ref. 22) is also worth noticing.

It is easy to see that after a straightforward algebraic manipulation Eq. \eqref{AveX} can be recast in the following simple form:
\begin{align}
\langle x_\alpha \rangle(z) =  w_\parallel {\langle x_\alpha \rangle}_\parallel(z) + w_\perp {\langle x_\alpha \rangle}_\perp (z), \label{AveX2}
\end{align}
where the \emph{weight} coefficient $ w_\parallel$ is defined as:
\begin{align}
 w_\parallel = & \; \frac{\displaystyle{\int  I_\parallel  \; d x \, d y}}{\displaystyle{\int  \left( I_\parallel + I_\perp \right) \; d x \, d y} } \nonumber \\
= & \; \frac{\displaystyle{\left|a_{\, \parallel} \, r_{\parallel} \right|^2}}{\displaystyle{\left|a_{\, \parallel} \, r_{\parallel} \right|^2 + \left|a_{\perp} \,
r_{\perp} \right|^2 } },   \label{WeightPar}
\end{align}
and the \emph{relative} shift as:
\begin{align}
\langle x_\alpha \rangle_\parallel(z) = & \; \frac{\displaystyle{\int  x_\alpha \left|f(-x + X_{ \parallel} ,y - Y_{\, \parallel},z) \right|^2  \; d x \, d y}}{\displaystyle{\int  \left|f(-x + X_{ \parallel} ,y - Y_{\, \parallel},z) \right|^2 \; d x \, d y} } . \label{RelShiftX}
\end{align}
The corresponding formulae for $ w_\perp$ and ${\langle x_\alpha \rangle}_\perp(z)$ can be obtained from Eqs. (\ref{WeightPar}-\ref{RelShiftX}) by exchanging the subscripts ``$\parallel$'' and ``$\perp$'' everywhere. Equation \eqref{WeightPar} has been obtained by exploiting the fact that since the double integration extends over the entire $xy$ plane, one has:
\begin{align}
\int  \bigl|f(-x \, +  \bigr. &  \, \bigl.  X_{ \parallel} ,y - Y_{\, \parallel},z) \bigr|^2 \; d x \, d y =  \nonumber \\
& \,   \int  \bigl|f(-x + X_{\perp} ,y - Y_{\perp},z) \bigr|^2 \; d x \, d y . \label{Assumpt1}
\end{align}
Equations \eqref{WeightPar} and \eqref{RelShiftX} have a simple physical interpretation. The non-negative weight coefficients $w_\parallel$ and $w_\perp$ give the fraction of the reflected light which is linearly polarized in the directions parallel and perpendicular to the plane of incidence, respectively. The relative shifts $\langle  x_\alpha \rangle_\parallel(z)$   and $\langle x_\alpha \rangle_\perp(z)$ are the first moments of the distributions $I_\parallel$ and $I_\perp$, and represent the displacements of the intensity distributions that would be detected behind a polarizer oriented along $\hat{\mathbf{x}}_r$ and $\hat{\mathbf{y}}_r$, respectively, at distance $z$ from the origin.

At this point,  conventional calculations require substitution of Eqs. \eqref{ComplexShifts} into Eq. \eqref{RelShiftX} and subsequent straightforward integration, either analytical for a few simple cases, or numerical for the more complicated ones. \cite{BaidaEtAl} However, as announced in the Introduction, such procedure leads to a confusion,  in determining GH and IF shifts, between the role played by the \emph{functional} shape $f(x,y,z)$ of the beam on one side, and the role played by the beam-shape independent shifts $X_{\, \parallel}, Y_{\, \parallel}$ and  $X_\perp , Y_\perp$ on the other side. This issue will be clarified in  Sec. III and practically illustrated in Sec. IV.
\section{Functional shifts}
In order to elucidate the problem posed at the end of the previous section, a seemingly   more abstract issue is illustrated  here to the reader in some detail: Given an \emph{arbitrary} complex function $f(x,y,z)$ square integrable over the whole $xy$ plane, calculate the first moments $\langle x \rangle$ and  $\langle y \rangle$ of the shifted distribution
\begin{align}
F(x,y,z) = \left|f (x-X, y - Y,z)\right|^2,
\end{align}
where $X,Y \in \mathbb{C}$ and  $|X|,|Y| \ll 1$. \cite{Note2}
In other words one is required to  calculate the following quantities:
\begin{subequations}\label{t150}
\begin{align}
\langle x \rangle = &  \; \frac{\displaystyle{\int x F(x,y,z) \, d x d y}}{\displaystyle{\int  F(x,y,z) \, d x d y }}, \label{t150a} \\
\langle y \rangle = &  \; \frac{\displaystyle{\int y F(x,y,z) \, d x d y}}{\displaystyle{\int  F(x,y,z) \, d x d y }} .\label{t150b}
\end{align}
\end{subequations}
The complex shifts  $X$ and $Y$ of the argument of the function are written as:
\begin{subequations}\label{t160}
\begin{align}
X  = &  \; X_1 + i \, X_2, \label{t160a} \\
Y  = &  \; Y_1 + i \, Y_2, \label{t160b}
\end{align}
\end{subequations}
where $X_n, Y_n \in \mathbb{R}$, with $n \in \{1,2 \}$. A first-order Taylor expansion about $X=0=Y$ furnishes:
\begin{align}\label{t170}
F(x,y,z) \cong & \; \abs{f(x,y,z)}^2 \nonumber \\
& \; - 2 \left[X_1 \, \text{Re}\left( f^* \frac{\partial f}{\partial x}\right) - X_2  \, \text{Im}\left( f^* \frac{\partial f}{\partial x}\right)\right] \nonumber \\
& \; - 2 \left[Y_1 \,  \text{Re}\left( f^* \frac{\partial f}{\partial y}\right) - Y_2  \, \text{Im}\left( f^* \frac{\partial f}{\partial y}\right)\right],
\end{align}
where hereafter $f$ is a shorthand for $f(x,y,z)$, not to be confused with $f (x-X, y - Y,z)$.  As usual, ``$\text{Re}$'' and ``$\text{Im}$'' denotes, respectively, the real and the imaginary part of a complex number: $\text{Re}(a+ib) = a$ and $\text{Im}(a+ib) = b$, with $a,b \in \mathbb{R}$.
Without loss of generality, we assume that the function $f$ drops to zero at infinity and that both the $x$- and $y$-moment of the distribution $|f|^2$ are zero, namely:
\begin{subequations}\label{t180}
\begin{align}
\abs{f(\pm \infty,y,z)}^2 = \; & 0 =\abs{f(x,\pm \infty,z)}^2 , \label{t180a} \\
\int x \abs{f(x,y,z)}^2\, d x d y  = \; & 0 = \int y \abs{f(x,y,z)}^2 \, d x d y. \label{t180b}
\end{align}
\end{subequations}
By using integration by parts, it is not difficult to show that, up to first order terms in $X_n, Y_n$, one obtains after substitution of Eq. (\ref{t170}) into Eqs. (\ref{t150}):
\begin{subequations}\label{t190}
\begin{align}
\langle x \rangle \cong X_1 + 2 &  \;   \text{Im}\left[ X_2 \,\frac{\displaystyle{ \int f^* \left(x \, \frac{\partial}{\partial x}\right) f \, d x d y}}{\displaystyle{\int  \abs{f}^2 \, d x d y }}  \right. \nonumber \\
 &  \phantom{aaaa} \left. + Y_2 \, \frac{\displaystyle{\int f^* \left(x \, \frac{\partial}{\partial y}\right) f \, d x d y}}{\displaystyle{\int  \abs{f}^2 \, d x d y }}\right], \label{t190a} \\
\langle y \rangle \cong   Y_1 + 2 &  \; \text{Im} \left[  Y_2 \,\frac{\displaystyle{ \int f^* \left(y \, \frac{\partial}{\partial y}\right) f \, d x d y}}{\displaystyle{\int  \abs{f}^2 \, d x d y }} \right. \nonumber \\
 &  \phantom{aaaa} \left.  + X_2 \, \frac{\displaystyle{\int f^* \left(y \, \frac{\partial}{\partial x}\right) f \, d x d y}}{\displaystyle{\int  \abs{f}^2 \, d x d y }}\right] .\label{t190b}
\end{align}
\end{subequations}
One should  appreciate the fact that these formulae are \emph{perfectly general}, as they hold their validity for any square-integrable function $f(x,y,z)$ fulfilling conditions (\ref{t180}).

The meaning of Eqs. \eqref{t190} becomes crystal clear when they are put in the following suggestive matrix form:
\begin{align}\label{matrixA}
\left[
\begin{array}{c}
  \langle x \rangle  \\
  \langle y \rangle
\end{array}
\right] \cong \left[
\begin{array}{c}
  X_1  \\
  Y_1
\end{array}
\right] +
\displaystyle{\left[
  \begin{array}{cc}
    a_{11} & a_{12} \\
    a_{21} & a_{22} \\
  \end{array}
\right]}
\left[
\begin{array}{c}
  X_2  \\
  Y_2
\end{array}
\right],
\end{align}
where the elements $a_{\alpha \beta}$ of the matrix $A = [a_{\alpha \beta}]$, with $\alpha, \beta =1,2,$ have been defined as:
\begin{align}\label{matrixAelements}
a_{\alpha \beta} = 2 \,\frac{\displaystyle{ \text{Im}\int f^* \left(x_\alpha \, \frac{\partial}{\partial x_\beta}\right) f \, d x d y}}{\displaystyle{\int  \left|f \right|^2 \, d x d y }}.
\end{align}
The first term in the right side of Eq. \eqref{matrixA} simply states a trivial fact: If one shifts a distribution, initially centered in the origin of the $xy$ plane, by the real amounts $X_1$ and $Y_1$  in the $x$ and $y$ directions, respectively, then the shifted distribution will be centered in $P = (X_1,Y_1)$.

Much more interesting is the second term. The diagonal elements of the $2 \times 2$ matrix $A$ connect the imaginary parts of the complex shifts $X$ and $Y$  with the real displacements of the distribution. Since from Eq. \eqref{matrixAelements} it trivially follows that $A$ must be a function of $z$ solely, this means that $X_2$ and $Y_2$ must somehow describe the propagation properties of the shifts $\langle x \rangle$ and $\langle y \rangle$, respectively. It will be explicitly shown in Sec. IV that this is indeed the case for functions $f$ describing paraxial beams of light, and that $a_{11}$ and $a_{22}$ are responsible for the \emph{angular} GH and IF shifts, respectively.

The non-diagonal matrix element $a_{12}$ ($a_{21}$) \emph{mixes} the real part of the shift $X$ ($Y$) in the $x$ ($y$) direction with the imaginary part of the shift $Y$ ($X$)  in the $y$ ($x$) direction.  The existence of these nonzero off-diagonal matrix elements is a remarkable \emph{universal} feature of shifted distributions. In fact, although the exact amount of mixing, namely the precise values of  $a_{12}$ and $a_{21}$, depends on the  functional shape of $f$, its existence is largely independent from the form of $f$.
This seemingly purely mathematical property of distributions has actually profound consequences on the physics of beam-shift phenomena. In fact,  such  a mixing  has been actually predicted \cite{Fedoseyev01,Bliokh:09} and experimentally observed \cite{AielloPra2010} for vortex beams.
Thus, for a better understanding of this issue,  in the following Sec. IV the perfectly general mathematical analysis of functional shifts given above will be specialized to the physical case of vortex beams, namely beam carrying OAM.
\section{Example: Functional shifts of vortex beams}
The goal of this section is to illustrate the use of the formula \eqref{matrixA} by calculating the coordinates of the centroid of a reflected \emph{vortex}  beam. According to Eq. \eqref{AnaSigRef} the shifted beam is described by one or two functions of the form $f(-x + X, y - Y, z)$, with
\begin{align}\label{t200}
f(x,y,z) = e^{i m \phi} g(r,z),
\end{align}
where $m$ is a real number, not necessarily an integer one, and $x =  r  \cos \phi, \; y =   r  \sin \phi$
are the polar coordinates in the $xy$ plane. Equation \eqref{t200}, with integer $m$, furnishes a suitable representation of optical beams with $m$ units of OAM along the direction of propagation $z$. For example, the well-known Laguerre-Gauss paraxial modes $u^\text{LG}_{m p}(\phi,r,z)$ of the electromagnetic field,  have the functional form \eqref{t200} with
\begin{align}
g(r,z) = & \,   \left[ \frac{2 p!}{\pi  \left(p + \abs{m} \right)!} \right]^{1/2} \exp \left({  \frac{i}{2} \frac{r^2}{z - i L} }  \right) \nonumber \\
 & \, \times \exp \left[ {-i \left(2 p + \abs{m} + 1 \right)\arctan \left( \frac{z}{L} \right) } \right]  \nonumber \\
& \, \times  \frac{1}{w(z)}L^{\abs{m}}_p \left( \frac{2 \, r^2}{w^2(z)} \right) \left( \frac{\sqrt{2} \, r  }{w(z)}\right)^{|m|} ,
\label{LGmodes}
\end{align}
where $w_0$ is the beam waist, $L = k^2 w_0^2/2$ is the Rayleigh range, $w(z) = k w_0 \sqrt{1 + {z^2}/{L^2}}$
is the \emph{running} beam waist, and $L^{\abs{m}}_p \left( x \right)$ denotes the generalized Laguerre polynomial \cite{GandR} of order $(m,p)$,
with $p = 0, 1,2,\dots$ and $m = 0, \pm1, \pm 2, \dots$. The first few Laguerre-Gauss modes intensity patterns are pictured in Fig. 3 below.
%
%
\begin{figure}[h!]
\begin{center}
\includegraphics[angle=0,width=7.5truecm]{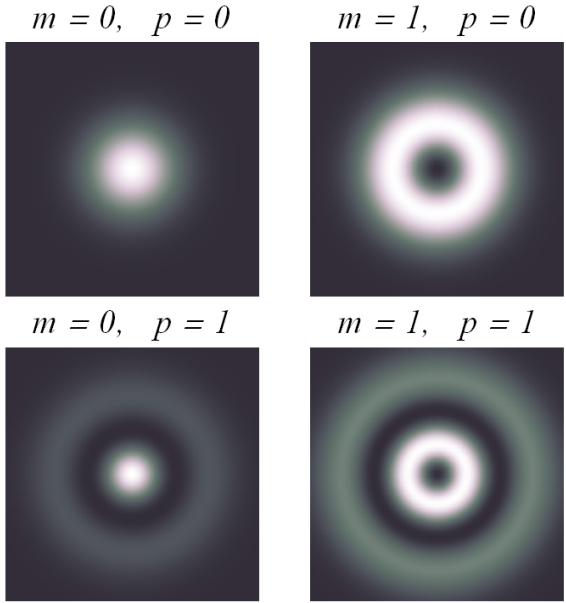}
\caption{\label{fig:3} Some low-order Laguerre-Gauss modes  intensity patterns $\left| u^\text{LG}_{m p}(\phi,r,z) \right|^2$.}
\end{center}
\end{figure}
%
%

For the sake of simplicity, hereafter  $X,Y$ denote either $X_\parallel, Y_\parallel$ or $X_\perp, Y_\perp$, we leave it unspecified. Correspondingly,  for example,  $\langle x \rangle$ denotes either $\langle x \rangle_\parallel$ or $\langle x \rangle_\perp$.
Thus, from Eqs.  \eqref{matrixAelements}-\eqref{t200}, we can obtain the matrix $A$ by calculating the following integrals:
\begin{align}
\int \abs{f(-x,y,z)}^2 \, dx dy = & \; 2 \pi \int \abs{g(r,z)}^2 r dr, \label{t220}
\end{align}
and
\begin{subequations}\label{t230}
\begin{align}
\int f^* \left( x \frac{\partial}{\partial x}\right) f \, dx dy  = \; & \pi \int_0^\infty g^*(r,z)\left( r \frac{\partial}{\partial r}\right) g(r,z) \, r dr \nonumber \\
\equiv \; &  N , \label{t230a} \\
\int f^* \left( x \frac{\partial}{\partial y}\right) f \, dx dy  =  \; & -i \pi m \int_0^\infty \abs{g(r,z)}^2\, r dr \nonumber \\
\equiv \; & i M \label{t230b}
\end{align}
\end{subequations}
where $f$ now is a shorthand for $f(-x,y,z)$.
The last remaining two integrals are not independent from the previous ones, being:
\begin{subequations}\label{t235}
\begin{align}
\int f^* \left( y \frac{\partial}{\partial y}\right) f \, dx dy  =  \; & N , \label{t230c} \\
\int f^* \left( y \frac{\partial}{\partial x}\right) f \, dx dy  =  \; & - i M .\label{t230d}
\end{align}
\end{subequations}
Finally, substitution of Eqs. (\ref{t230}) into Eq. (\ref{t190}) yields:
\begin{align}\label{matrixA2}
\left[
\begin{array}{c}
  \langle x \rangle  \\
  \langle y \rangle
\end{array}
\right] \cong \left[
\begin{array}{c}
  X_1  \\
  Y_1
\end{array}
\right] +
\displaystyle{\left[
  \begin{array}{cc}
    \zeta(z) & -m \\
    m & \zeta(z) \\
  \end{array}
\right]}
\left[
\begin{array}{c}
  X_2  \\
  Y_2
\end{array}
\right],
\end{align}
where
\begin{align}
\zeta (z) = \frac{\displaystyle{\text{Im}\int_0^\infty g^*(r,z)\left( r \frac{\partial}{\partial r}\right) g(r,z) \, r dr}}{\displaystyle{\int_0^\infty \abs{g(r,z)}^2 \, r dr}}
\label{t250} ,
\end{align}
determines  how the shifted distribution ``propagates'' along $z$. In physical terms, the real function $\zeta(z)$ generates the angular GH and IF shifts. This may be easily seen by deriving both sides of Eq. \eqref{matrixA2} with respect to $z$ and comparing such result with the right sides of  Eqs. \eqref{GHandIFShifts}:
\begin{align}\label{matrixA3}
\left[
\begin{array}{c}
 \Theta_\text{GH}  \\
 \Theta_\text{IF}
\end{array}
\right] \equiv
\frac{\partial}{\partial z} \left[
\begin{array}{c}
  \langle x \rangle  \\
  \langle y \rangle
\end{array}
\right] \cong  \frac{\partial \zeta(z)}{\partial z}
\left[
\begin{array}{c}
  X_2  \\
  Y_2
\end{array}
\right].
\end{align}

Of course, the precise form of $\zeta(z)$ can be given only when $g(r,z)$ is precisely specified. Thus, for example, for
 a Laguerre-Gauss paraxial beam $g(r,z)$ of the form \eqref{LGmodes},  one obtains from Eq. \eqref{t250}:
\begin{align}
\zeta (z) = \left( 1 + \abs{m} \right) \frac{z}{L}\label{t260},
\end{align}
where, for the sake of clarity, the value $p=0$ has been chosen.
When this expression is substituted into Eq. \eqref{matrixA2} the following result is obtained:
\begin{subequations}\label{t270}
\begin{align}
\langle x \rangle \cong &  \; X_1 - m Y_2 + \frac{z}{L} \left( 1 + \abs{m} \right)  X_2, \label{t270a} \\
\langle y \rangle \cong &  \; Y_1 + m X_2 + \frac{z}{L} \left( 1 + \abs{m} \right)  Y_2. \label{t270b}
\end{align}
\end{subequations}

If $m=0$ the  Laguerre-Gauss beam reduces to a  fundamental Gaussian beam, and  Eqs. \eqref{t270} become:
\begin{subequations}\label{t280}
\begin{align}
\langle x \rangle^{0} \cong &  \; X_1  + \frac{z}{L}   X_2, \label{t280a} \\
\langle y \rangle^{0} \cong &  \; Y_1  + \frac{z}{L} Y_2 , \label{t280b}
\end{align}
\end{subequations}
where hereafter the superscript ``$0$'' denotes quantities relative to a fundamental Gaussian beam. This leads to the unambiguous identification
\begin{subequations}\label{t290}
\begin{align}
X_1 & \;= \Delta_\text{GH}^0, & X_2   = & \; L \Theta_\text{GH}^0, \label{t290a} \\
Y_1 & \; = \Delta_\text{IF}^0, & Y_2  = & \; L \Theta_\text{IF}^0, \label{t290b}
\end{align}
\end{subequations}
where Eqs. \eqref{GHandIFShifts} have been used.
Finally, by using Eqs. \eqref{t290} into  Eqs. \eqref{t270}, one easily obtains:
\begin{subequations}\label{t300}
\begin{align}
\langle x \rangle \cong &  \; \Delta_\text{GH}^0 + m \left(  L  \Theta_\text{IF}^0 \right) + z \left( 1 + \abs{m} \right)  \Theta_\text{GH}^0, \label{t300a} \\
\langle y \rangle \cong &  \; \Delta_\text{IF}^0 - m \left(  L   \Theta_\text{GH}^0 \right) + z \left( 1 + \abs{m} \right)  \Theta_\text{IF}^0. \label{t300b}
\end{align}
\end{subequations}
This outcome coincides, except for a few minor discrepancies due to the use of a different notation, with the well-known results for vortex beams. \cite{Fedoseyev01,Bliokh:09,AielloPra2010} From Eqs. \eqref{t300} it appears that OAM of light has
a twofold effect upon GH and IF shifts. The first one is an \emph{algebraic}  mixing proportional to $m$  between spatial GH (IF) and angular IF (GH). The second effect is an \emph{increase} of the angular shift by the amount $|m|$. As revealed by the approach presented in this article, these features are universal for Laguerre-Gauss beams, namely they are independent upon both the beam polarization and the material composition of the reflecting surface.

Thus, Eqs. \eqref{t300} and \eqref{matrixA}  answer the two questions posed in the introduction, namely: \emph{a}) how the OAM of light affects GH and IF shifts, and: \emph{b}) to what extent these shifts are determined by the shape of the beam (see discussion after Eq. \eqref{matrixAelements}).
\section{Concluding remarks}
The main  result presented in this paper perhaps consists in the following recipe for the calculation of GH and IF shifts: Use the conventional theory to calculate the \emph{beam-shape independent} complex beam shifts $X_\parallel, Y_\parallel$ and $X_\perp, Y_\perp$, for your peculiar surface.   They depend upon both the polarization of the beam and the material composition of the reflecting surface, but are \emph{totally} independent from the shape of the beam. Then, by using Eq. \eqref{matrixAelements}, calculate the matrix $A$ for your specific beam. Finally, combine the previous results into Eq. \eqref{matrixA} to obtain the desired shifts.

Apart from its great practical usefulness, this new approach has the important conceptual advantage over the traditional beam-shift computational methods, that it permits to isolate in a completely unambiguous manner the beam-shape dependent contributions to the shifts. This separation is relevant whenever one wants to understand how the degrees of freedom carried by the  wavefront of the beam (as, e.g., the OAM of the beam), enter in the numerous beam-shift physical phenomena.
\begin{acknowledgments}
It is a pleasure for the author to thank Han Woerdman, Vladimir Fedoseyev, Konstantin Bliokh, Mark Dennis, Sumant Oemrawsingh and Michele Merano for many stimulating discussions. Many thanks also to Marco Ornigotti, Falk T\"{o}ppel, Annemarie Holleczek, Jan Korger and Giuliano d'Auria for careful proofreading of the manuscript.
\end{acknowledgments}
\vfill

\vfill

\begin{thebibliography}{23}
\bibitem{Note}
Here, the minus sign in the $x$-dependence of the shifted distribution should be noticed. It is not difficult to see that this is due to the parity inversion caused by reflection as seen from the reflected-beam reference frame. The elementary proof of this statement is left to the reader as exercise.
\end{thebibliography}
\end{document}